%% file: main.tex
\documentclass{article}
\input{sections/preamble}
\begin{document}

\maketitle

\let\thefootnote\relax\footnotetext[1]{${}^*$This paper is based on our invited article in TSUBAME
  e-Science Journal (ESJ)\cite{ESJ2017}. ESJ is a non-academic and non-peer-reviewed
  newsletter from Global Scientific Information Center, Tokyo
  Institute of Technology.}

\input{sections/abstract}
\input{sections/introduction}
\input{sections/fundamentals}

\input{sections/impl}
\input{sections/eval}
\input{sections/conclusion}

\input{sections/acknowledgements}

\small
\bibliographystyle{abbrv}
\bibliography{main}

\end{document}

%% file: sections/preamble.tex
\PassOptionsToPackage{square, numbers, compress,sort}{natbib}
\usepackage[final]{nips_2017}
\usepackage{graphicx}

\usepackage[T1]{fontenc}
\usepackage{booktabs}
\usepackage{amsfonts}
\usepackage{nicefrac}
\usepackage{microtype}
\usepackage[utf8]{inputenc}
\usepackage{color}
\usepackage{graphicx}
\usepackage{algorithm,algorithmic}
\usepackage{hyperref}
\usepackage{url}
\usepackage{amsfonts}
\usepackage{amsmath}
\usepackage{booktabs}
\usepackage{wrapfig}
\usepackage{listings}
\usepackage{subfigure}
\usepackage[font=small]{caption}

\title{ChainerMN: Scalable Distributed Deep Learning Framework \footnotemark}

\author{
  Takuya Akiba \\
  Preferred Networks, Inc. \\
  \texttt{akiba@preferred.jp}
  \And
  Keisuke Fukuda \\
  Preferred Networks, Inc. \\
  \texttt{kfukuda@preferred.jp}
  \And
  Shuji Suzuki \\
  Preferred Networks, Inc. \\
  \texttt{ssuzuki@preferred.jp}
}

%% file: sections/abstract.tex
\begin{abstract}
  One of the keys for deep learning to have made a breakthrough in
  various fields was to utilize high computing powers centering around
  GPUs. Enabling the use of further computing abilities by distributed
  processing is essential not only to make the deep learning bigger
  and faster but also to tackle unsolved challenges. We present the
  design, implementation, and evaluation of ChainerMN, the distributed
  deep learning framework we have developed. We demonstrate that
  ChainerMN can scale the learning process of the ResNet-50 model to the
  ImageNet dataset up to 128 GPUs with the parallel efficiency
  of 90\%.
\end{abstract}

%% file: sections/introduction.tex
\section{Introduction}

It has turned out that deep learning achieves far better predicting
performance than existing methods in image recognition, natural
language processing, speech recognition and many other fields where
machine learning is being applied. The basic technology of neural
networks used in deep learning has a long history dating back to the
1950’s. As we entered the 2010’s, the neural network technology with
its long history has made the breakthrough as ``deep learning'' as
described above because it is thought to have successfully combined
all the advances of algorithms, large-scale data, and high computing
powers. Even today, it would be difficult to achieve an outstanding
predicting performance by deep learning if one of the three lacks. In
this article, we focus on one of the three pillars supporting deep
learning: computing performance.

It has become a standard approach to use highly efficient GPUs for
training in many deep learning tasks. Nevertheless, the training
process is still time-consuming even with the latest GPUs because
models have also grown massive and complex. For example, training
Resnet-50~\cite{He2016} for the ImageNet dataset~\cite{imagenet2009}
typically takes as long as one week with a single GPU. Taking a long
time on training means you have a limited number of times to do trial
and error for models and parameters needed to achieve high accuracy,
making it difficult to produce a good predicting performance. It also
means there is a limit to the usable data size. Thus, using multiple
GPUs in parallel is crucial in accelerating calculation.

We introduce ChainerMN, an add-on package to Chainer~\cite{Tokui2015},
a programming framework for deep learning applications written in
Python, to provide a distributed learning capability. In the course of
developing ChainerMN, we took the following features into
consideration:

\begin{itemize}
\item \textbf{Flexibility:} Chainer is a flexible framework based on
  its Define-by-Run approach and ChainerMN is designed not to ruin the
  flexibility aspect. This allows for easy distributed learning even
  in complex use cases such as dynamic neural networks, generative
  adversarial networks, and reinforced deep learning.
\item \textbf{High performance:} We selected technologies assuming
  practical workloads in deep learning from the very beginning of
  designing ChainerMN as well as exercised ingenuity with respect to
  implementation so that hardware performance is fully utilized.
\end{itemize}

The rest of the paper is organized as follows. First, we explain the
basic elements of distributed deep learning, followed by the design
and implementation of ChainerMN. Finally, we will present the results
of our evaluation experiment and related work.

%% file: sections/fundamentals.tex
\section{Preliminaries}
\label{sec:preliminaries}

\subsection{Basics of Deep Learning}

We can express the prediction by neural networks against input data
$x$ as $f(x;\theta)$ where $\theta$ is a parameter for neural networks.
Learning in neural networks using backpropagation and stochastic
gradient descent or its variations is an iterative algorithm. Each
iteration is composed of the following three steps: forward
computation, backward computation, and optimization.

In the forward-computation step, first, the prediction $f(x;\theta)$
is calculated against an input data point $x$. Then, the loss is
calculated to represent the difference from the correct output
for. Here, the cross entropy and other indicators may be used.

In the backward-computation step, $g=\frac{\delta E}{\delta \theta}$ ,
the gradient of the parameter $\theta$ in the direction of decreasing
the loss $E$, is calculated. Gradients for all parameters are
calculated using the chain rule while going backward from the output
layer to the input layer.

In the optimize step, the parameter $\theta$ is updated using the
gradient $g$ . The simplest rule is to update $\theta$ to $\theta -
\mu g$ where $\mu$ is a parameter called a learning rate.

In practice, instead of using a single training example in an
iteration, the forward and backward calculations are performed
simultaneously against multiple training examples and optimization is
executed using the average of gradients against all the examples. The
input examples used in an iteration is called a minibatch while its
size is called a batch size. A typical batch size ranges from several
tens to several hundred.

Please note that the above description is based on a standard
supervised learning. Nonetheless, in case that neural networks are
applied to other algorithms such as unsupervised learning and
semi-supervised learning, the parallelizing method we will explain
below is applicable and ChainerMN is also usable.

\subsection{Data Parallel and Model Parallel approaches}

\begin{figure}[ht]
  \begin{center}
    \begin{minipage}[b]{0.45\hsize}
      \centering
      \subfigure[Data parallelism.]{
        \includegraphics[clip,height=85pt]{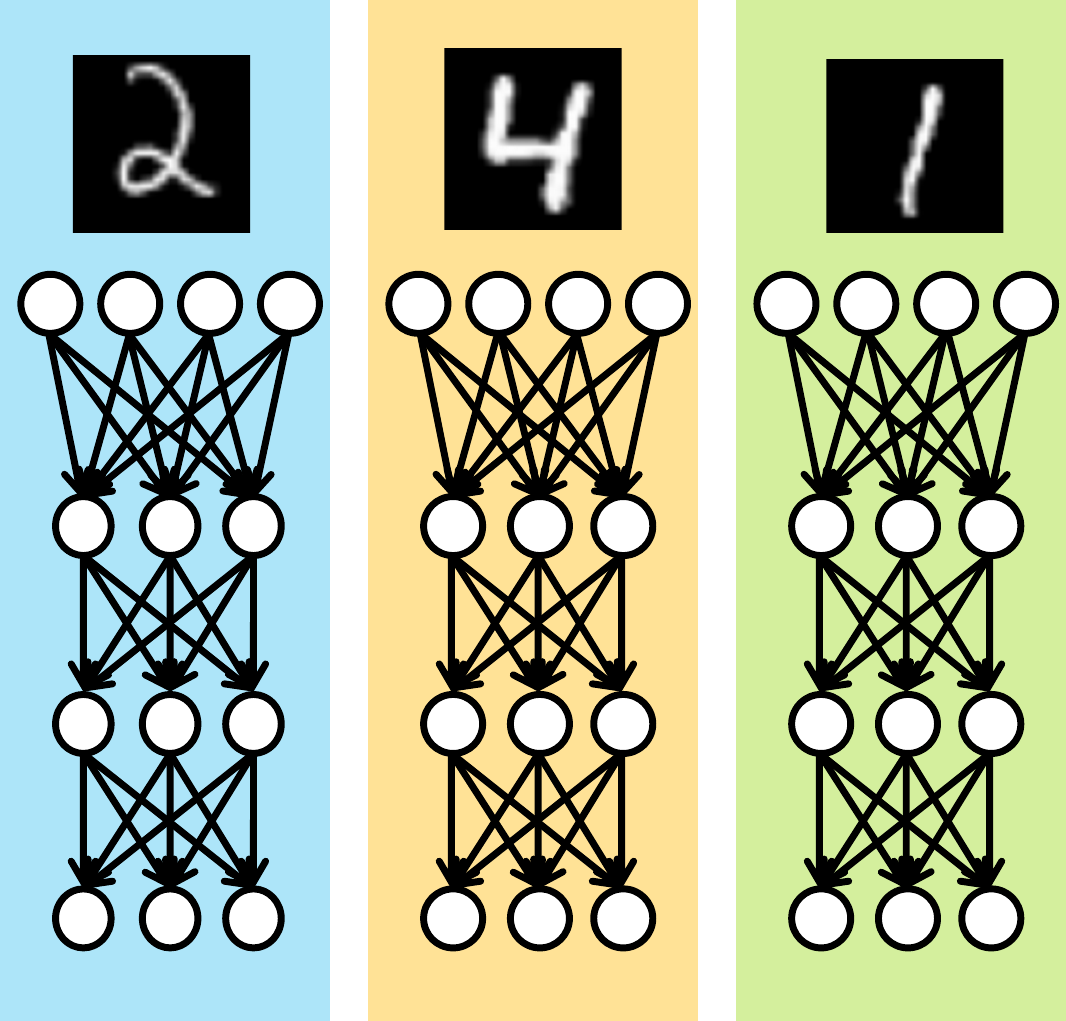}}
      \hspace{0.25em}
      \subfigure[Model parallelism.]{
        \hspace{0.8em}
        \includegraphics[clip,height=85pt]{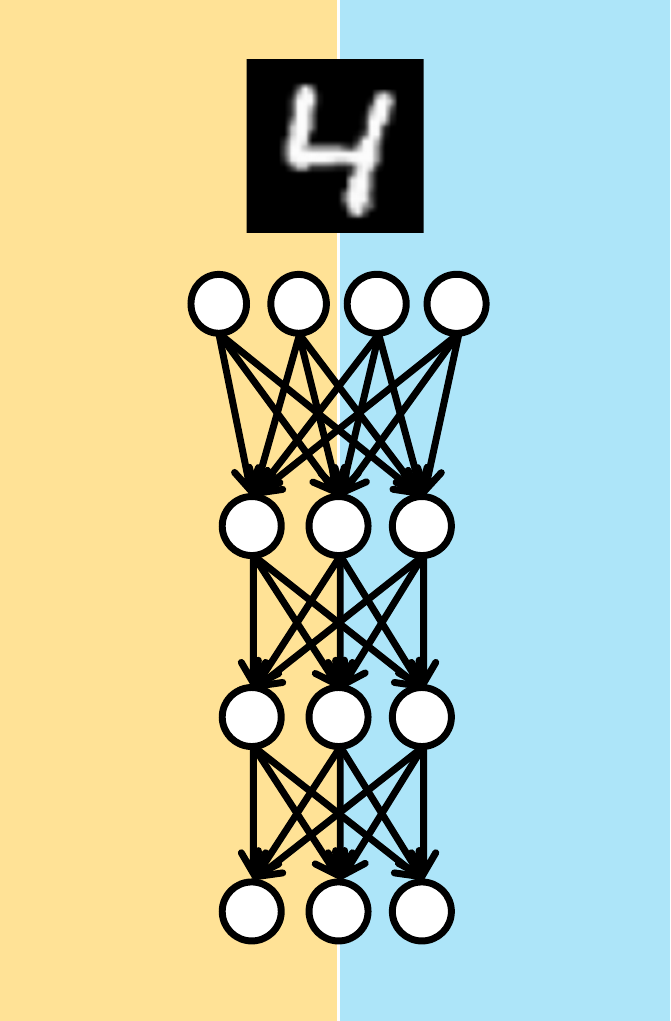}
        \hspace{0.8em}}
      \caption{Data parallel and model parallel approaches.}
      \label{fig:data-par-model-par}
    \end{minipage}\hspace{1em}\begin{minipage}[b]{0.45 \hsize}
      \centering
      \includegraphics[clip,width=0.95\columnwidth]{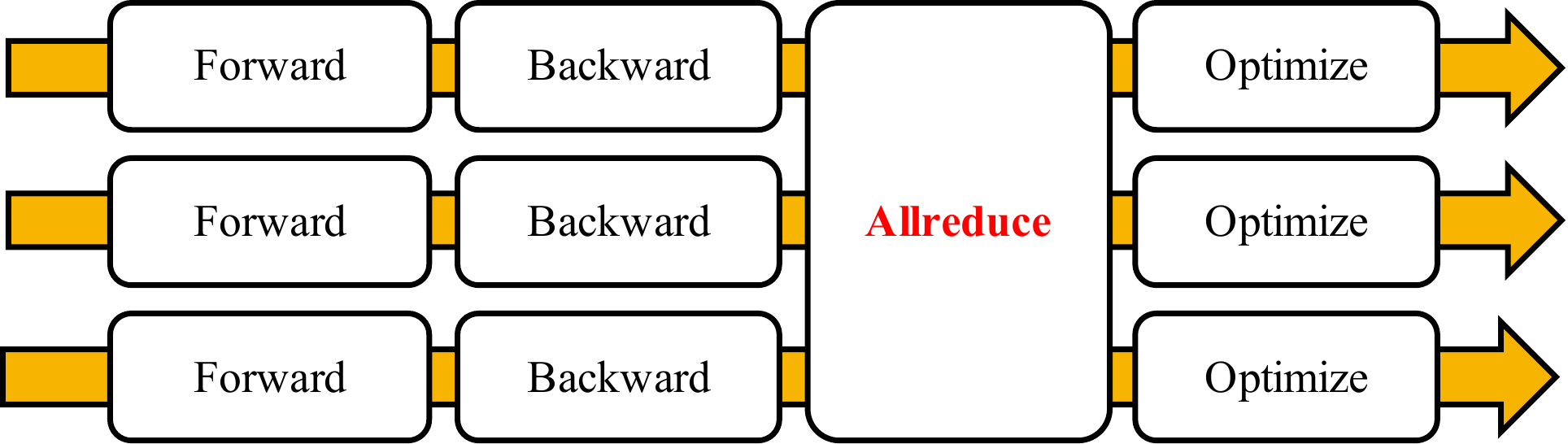}
      \vspace{3em}
      \caption{The four steps that constitute an iteration of
        synchronous data-parallel approach.}
      \label{fig:four_steps}
    \end{minipage}
  \end{center}
\end{figure}

There are two major approaches to parallelize training by
distributed processing: data parallel and model parallel. In the
data-parallel approach, each worker has a model replica and calculate
gradients of different minibatches. Workers use these gradients to
update the model collaboratively. In the model parallel approach, each
worker has a portion of the model and work in cooperation with others
to do the calculation for one minibatch.
Figure~\ref{fig:data-par-model-par} shows the difference between the
two approaches.

The model-parallel approach was actively used in the days when GPU
memory was small. At present, the model parallel is rarely used in its
basic form as the data parallel approach is being used. In the
meantime, some issues with the data paralleled approach have surfaced
while research on a new form of the model parallel is underway. The
model parallel and the data parallel can be used at the same time as
well.

\subsection{Synchronous vs.~Asynchronous}

In this subsection, we will focus on the data-parallel approach which
is commonly used now. The data-parallel approach is roughly divided
into synchronous and asynchronous types, and we explain about the
former first. Each iteration in synchronous, data-parallel deep
learning is composed of the following four steps: forward computation,
backward computation, \texttt{Allreduce} communication, and
optimization. Figure~\ref{fig:four_steps} illustrates the four steps.

This has an additional step \texttt{Allreduce} to the regular
iteration described earlier. In this step, workers communicate with
each other to find the average of gradients calculated by individual
workers and distribute the average. All workers update the model using
the gradient they have obtained through the communication. If we
define the batch size processed by each worker as $b$ and the number
of workers as $n$, the gradient obtained through communication is
equivalent to the gradient in the batch size $bn$. This means gradients
are calculated using more training data in one iteration, improving
the gradient quality and accelerating the learning process.

Asynchronous type, on the other hand, uses special workers called a
parameter server. The parameter server controls model
parameters. Normal workers send gradients to the parameter server once
the gradients are obtained by forward and backward calculations. The
parameter server receives and uses the gradients to update the
model. Workers receive new model parameters and calculate gradients
again.

%% file: sections/impl.tex
\section{Design and Implementation}

\subsection{Parallelization Approaches}

We discuss the design decision of ChainerMN in this section.  As we
discussed in section \ref{sec:preliminaries}, there are two major
parallelization approaches and two synchronization approaches.  We
adopt a synchronous and data parallel approach for ChainerMN.

We use the data parallel approach because existing deep learning
applications would easily be extensible and faster training process
through data parallel was highly expected. Data
parallelization is tantamount to increasing a minibatch size in a
typical deep learning application and has its advantage of being
applicable without having to make significant changes in algorithms and
codes of existing programs.

Whether the synchronous or asynchronous type is desirable is also a
nontrivial question since different kinds of strategies have been
taken in each implementation and results would vary depending on tasks
or settings. The paper \cite{Xinghao2017} shows experimental results
that the asynchronous type is less stable regarding convergence
whereas it is faster to converge in the synchronization. Also, we can
benefit from the optimized and proven group communication mechanism of
MPI, the de-facto standard communication library interface, while in
the asynchronous model the implementation scheme uses a parameter
server in general.

\subsection{Chainer}

Chainer is a framework with its Define-by-Run feature. Define-by-Run
is a model that takes advantage of the flexibility of script languages
where learning models and computational flows are defined at
runtime. A Define-and-Run approach, on the other hand, is a model that
pre-defines a structure of networks, after which data is input and
calculation are done. While potentially easier to optimize performance,
this approach is said to lack flexibility.

Chainer provides programming models that enable to define complex
neural networks flexibly or make modifications during runtime thanks
to its Define-by-Run approach. This lets researchers and engineers
work on new models or complex models through trial and error with ease
and therefore is suitable for research and development of machine
learning.  Upon development, we carefully designed the ChainerMN API
with the objective of making it easily portable from existing Chainer
programs without putting limitations on the flexibility of Chainer.

\subsection{API Design}
We describe the design of library interface of ChainerMN by describing
minimal steps to extend an existing deep learning program written in
Chainer to support distributed execution using ChainerMN.

Listing~\ref{lst:chainermn} shows a simplified ChainerMN program of a
model to solve MNIST classification problem~\cite{mnist2009}. For a
complete program code, refer to ChainerMN's code
repository~\cite{chainermn}. There are three major steps: \textit{(1)}
add a communicator component, \textit{(2)} create and use
\texttt{mutli\_node\_optimizer}, and \textit{(3)} add code to
distribute a dataset.

A process of modifying an application starts from adding a
communication component called \texttt{Communicator} to existing
Chainer programs. A communicator is a central component of ChainerMN,
and it is designed after MPI's communicator concept and controls all
inter-process communication in ChainerMN program.

\texttt{mutli\_node\_optimizer} is the most important component in
ChainerMN.  It wraps Chainer's normal optimizer and exchanges the
gradient across processes using \texttt{Allreduce} operation before
optimizing the model. \texttt{multi\_node\_optimizer} behaves
identically as the original optimizer except for the communication, so
the extension is seamlessly integrated into Chainer's existing
\texttt{Trainer} ecosystem.

On top of this, basic porting can be done just by adding the
scattering step which distributes data for data parallel computations.
One needs to split the dataset into equal chunks and distribute them
over the processes. This operation is also known as \texttt{Scatter} in
MPI. Other parts, i.e.\texttt{Iterator},
\texttt{Updater}, and \texttt{Evaluator} do not need to be changed in
basic use cases. Because of this API design, it allows various Chainer
programs to be ported with minimal modifications while making the most
of the advantage given by Define-by-Run.

\begin{figure}[t]
\begin{lstlisting}[caption={Example of ChainerMN},label=lst:chainermn]
  model = L.Classifier(MLP(args.unit, 10))

  # Create a communicator
  comm = chainermn.create_communicator()

  # Distribute a dataset
  train = chainermn.scatter_dataset(train, comm, shuffle=True)

  # Create and use multi_node_optimizer
  optimizer = chainermn.create_multi_node_optimizer(
        chainer.optimizers.Adam(), comm)
        optimizer.setup(model)

  # Use Chainer's Trainer class to simplify
  # a forward-backward-optimization loop
  train_iter = chainer.iterators.SerialIterator(train, args.batchsize)
  updater = training.StandardUpdater(train_iter, optimizer, device=device)
  trainer = training.Trainer(updater, (args.epoch, 'epoch'), out=args.out)
\end{lstlisting}
\end{figure}

\subsection{Implementation and Performance Optimization}
The communication pattern of synchronous and data parallel deep
learning applications is relatively simple from the point of view of
HPC applications. Roughly speaking, the only major communication is
\texttt{Allreduce}, a process to exchange gradients which are training
and evaluation results. Auxiliary parts include \texttt{Scatter},
which arranges necessary data over distributed processes before
starting training.

As mentioned above, one of the design goals of ChainerMN is to achieve
high performance by leveraging existing and proven HPC technologies.
\texttt{Allreduce} is a component that especially requires speed
because it is called in every training iteration and needs to process
a large amount of data. We attempt to minimize the communication time
by using NCCL~\cite{nccl} library developed by NVIDIA. NCCL is a
highly-optimized communication library which provides a faster
\texttt{Allreduce} operation between NVIDIA GPUs within and across
nodes. 

%% file: sections/eval.tex
\section{Evaluation}

\subsection{Experimental Environment and Settings}
We conducted our experiments on our in-house cluster. It consists of
32 computing nodes.  Each node is equipped with two Intel Xeon CPUs
(E5-2623 v3, 3.00 GHz, four cores for each), 128 GB of main memory, and
four GeForce GTX TITAN X GPUs.  Thus, we used 128 GPUs in total.  The
nodes are interconnected by Mellanox Infiniband FDR 4X.  We used CUDA
version 8, Python version 3.5, Mvapich2 2.2 and Chainer version 1.2
running on Ubuntu 14.04 LTS.

To demonstrate the performance and scalability of ChainerMN, we use
ResNet-50~\cite{He2016} model and ImageNet~\cite{imagenet2009} dataset.
Since the dataset is large and the majority part of access is read, we
copied all the dataset to all computing nodes' local SSD in advance.

We used 32 as the batch size per GPU, which means 4096 for 128 GPUs.
One of the factors making distributed deep learning difficult is that
improving throughput does not necessarily mean better learning
efficiency. We note that the batch size 4096 is a healthy setting
where the learning efficiency and the resulting model accuracy are
maintained, as shown by Goyal et al.~\cite{Goyal2017}

\subsection{Scalability Result}

Figure~\ref{fig:scalability} shows the scalability of ChainerMN up to
128 GPUs.  In this figure, ChainerMN scales well up to 128 GPUs.
Table \ref{tab:par-eff} shows the relative runtimes over one-GPU
execution.  In this table, ChainerMN on 128 GPUs achieves 79 \% and
90 \% parallelization efficiency of the one-GPU and one-node (four
GPUs) executions, respectively.  It means that the parallelization
efficiency of ChainerMN on 128 GPUs is as high as the state-of-the-art
\cite{Goyal2017}.

\begin{minipage}{\textwidth}
  \vspace{1em}
  \begin{minipage}{0.45\textwidth}
    \centering
    \includegraphics[width=1.0\textwidth]{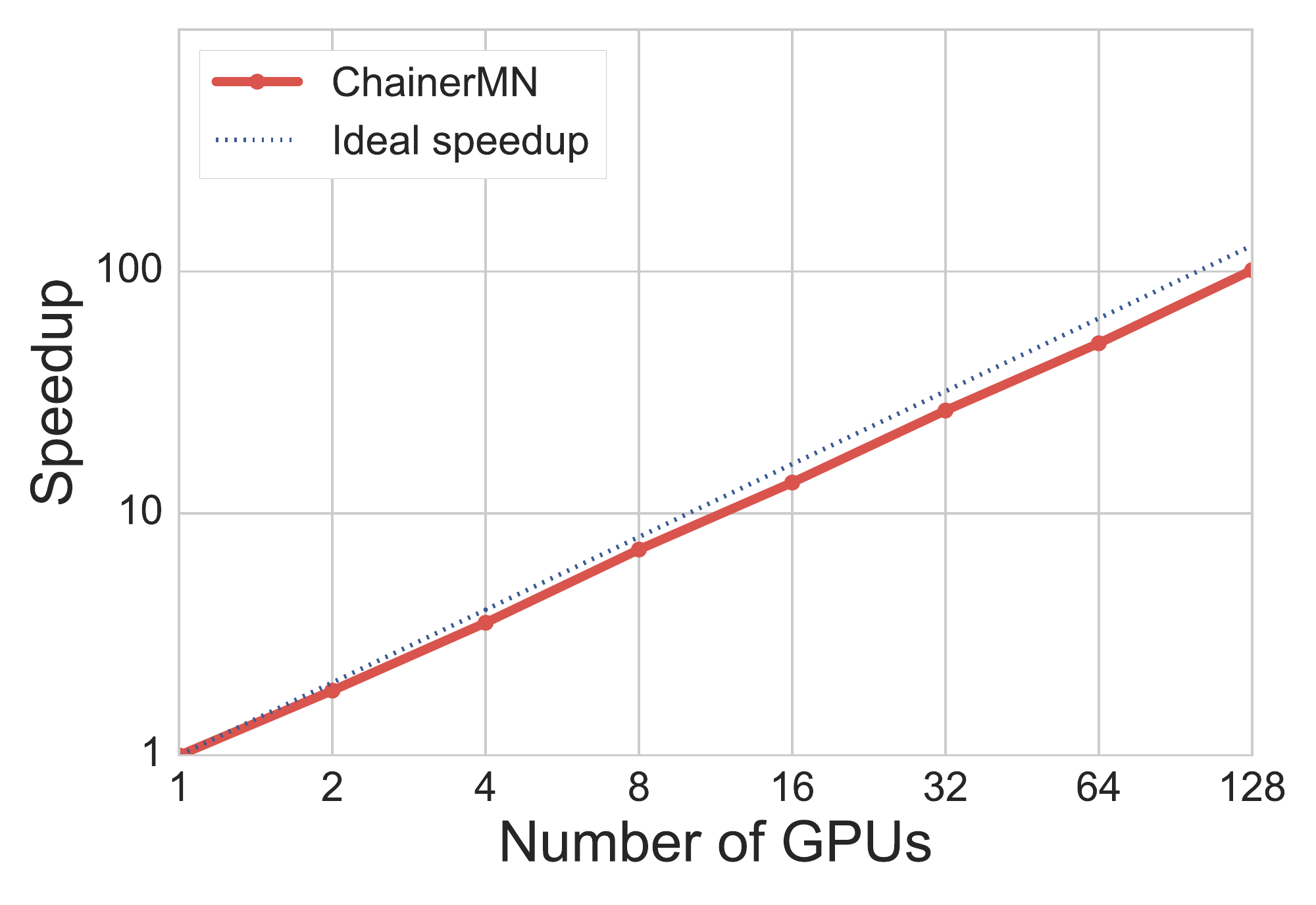}
    \captionof{figure}{Scalability of ChainerMN}
    \label{fig:scalability}
  \end{minipage}
  \hfill
  \begin{minipage}{0.55\textwidth}
  \centering
    \captionof{table}{Relative speed-up and parallelization efficiency}
    \small
  \label{tab:par-eff}
  \begin{tabular}{r|rr}  \toprule
    \#GPUs & Speed-up & Par. Eff. \\ \midrule
    1	       & 1.00    & 100.00\%  \\
    2        & 1.85    & 92.66\%   \\
    4        & 3.53    & 88.34\%   \\
    8        & 7.09    & 88.67\%   \\
    16       & 13.42   & 83.88\%   \\
    32       & 26.63   & 83.22\%   \\
    64       & 50.52   & 78.94\%   \\
    128      & 101.32  & 79.16\%   \\ \bottomrule
  \end{tabular}
  \end{minipage}
  \end{minipage}

%% file: sections/conclusion.tex
\section{Conclusions}

We have described the design and implementation of ChainerMN and
demonstrated its scalability. Chainer and ChainerMN are designed to
have both high flexibility and scalability with its primary object of
accelerating research and development in deep learning. We will
continue making improvements by tackling challenges such as model
parallel, overlapping communication and computation, asynchronous
computation among workers, optimized communication by compressed
gradients, and fault tolerance.

%% file: sections/acknowledgements.tex
\section*{Acknowledgements}
\small The authors thank K.~Ueno, T. Mitsuishi, and N.~Yoshifuji for
help on the development of ChainerMN.  We also thank T.~Sudo, Y.~Doi,
G.~Watanabe, R.~Okuta, and M.~Sakata for help for experiments.  We are
grateful to T.~Miyato and S.~Tokui for fruitful discussions as well.